\documentclass[aps,preprint]{revtex4}
\usepackage{amsfonts}
\usepackage{amssymb,epsfig}
\usepackage{latexsym}
\usepackage{amsmath}
\usepackage{graphicx}

\begin{document}

\title{Gauss-Bonnet dark energy Chaplygin Gas Model }
\author{A. Khodam-Mohammadi\footnote{\texttt{Email}: \text{Khodam@basu.ac.ir}}%
,~ E. Karimkhani\footnote{\texttt{Email}: \text{E.karimkhani91@basu.ac.ir}}%
~and~ A. Alaei} \affiliation{Department of Physics, Faculty of
Science, Bu-Ali Sina University, Hamedan 65178, Iran}

\begin{abstract}
The correspondence of the Gauss-Bonnet (GB) and its modification
(MGB) models of dark energy with the standard and generalized
Chaplygin gas-scalar field models (SCG and GCG) have been studied in
a flat universe. The exact solution of potentials and scalar fields,
which describe the accelerated expansion of the universe, are
reconstructed. According to the evolutionary behavior of the GB and
MGB models, the same form of dynamics of scalar field and potential
for different SCG and GCG models are derived. By calculating the
squared sound speed of the MGB, GB model as well as the SCG, GCG,
and investigating the GB-Chaplygin gas from the viewpoint of linear
perturbation theory, we find that the best results which is
consistent with the observation, may be appeared by considering the
MGB-GCG. Also we find out some bounds for parameters.
\end{abstract}

\maketitle


\section{\protect\bigskip Introduction}
Astrophysical data which is out coming from distant Ia supernova
\cite{A.J.Riess, S.Perlmutter, M. Hicken}, Large Scale Structure
(LSS) \cite{Tegmark, K. Abazajian} and Cosmic Microwave Background
(CMB)\cite{spergel, komatsu}, indicate that our universe undergoes
with an accelerating expansion. This kind of expansion may be arisen
by a mysterious energy component with negative pressure, so called,
dark energy (DE).

However in the last decades, other models based on modified gravity
($F(R), F(G), F(R,\phi,X), F(T)$,...) have been proposed that have
given another description of acceleration expansion of the universe.
In these models, many authors have showed that all models of DE can
be resolved by modifying the curvature term $R$ (Ricci scalar) of
Einstein-Hilbert action with another curvature scalars such as any
scalar function of $R$, Gauss-Bonnet term ($G$), torsion ($T$),
scalar-tensor (X,$\phi$) and etc. (details are in Ref. \cite{modG1,
modG2, modG3, modG4, modG5, modG6, modG7, modG8, modG9, modG10,
modG11, modG12, modG13, modG14} and references there in). Even, some
authors found that, the early inflation, the intermediate
decelerating expansion and late time acceleration expansion, could
be described together in one model
\cite{Chavanis}. \\
Lately, among many models of DE, dynamical models, which are
considering a time dependent component of energy density and
equation of state, have attracted a great deal of attention. Also,
among many dynamical models, ones that represented by a power series
of Hubble parameter and its derivative (\emph{i.e.} $\dot{H},
H\dot{H},H^{2},...$) have been interested \cite{sola1409, sola1412}.
Also authors in \cite{sola31, Perico, Basilakos13, bleem, Lima} have
shown that terms of the form $H^{3}$, $\dot{H}H^{2}$ and $H^{4}$ can
be important for studying of the early universe. Hence, it would not
be some thing strange to consider a DE density proportional to the
Gauss-Bonnet (GB) term which is invariant in 4-dimensional. Besides,
in geometrical meaning, the GB invariant has a valid dimension of
energy density \cite{GRANDA}. Also authors in \cite{kofinas, brown}
showed that a unification between early time inflation and late time
acceleration in a viable cosmology can be described by a coupling
between GB term and a time varying scalar field \cite{nojiri222}. \\
The other successful model of DE is Chaplygin Gas model. The
standard Chaplygin Gas model (SCG), first proposed by
\cite{chaplygin, kamens, bento}, regards as a perfect fluid which
plays a dual role in the history of the universe: it behaves as dark
matter in the first epoch of evolution of the universe and as a dark
energy at the late time. Unfortunately this model has some
inconsistency with observational data like SNIa, BAO, CMB
\cite{gorini, zhu, bento2}. So Generalized Chaplygin Gas (GCG)
\cite{bilic} and Modified Chaplygin Gas (MCG) models \cite{denbath,
Brown1, Cai1} have been introduced in order to establish a viable
cosmological model. It would be beneficial to study any relationship
between SCG model and its modification while DE density behaves like
GB invariant term as mentioned above.
In this paper we would show that it leads interesting cosmological implications.\\
As we would show in this paper, the EoS parameter of GB DE model on
its own does not give rise to phantom phase of the universe. Besides
in \cite{GRANDA}, author shows that presence of matter drastically
converts Friedmann equation into a nonlinear differential equation
which alters the behavior of the EoS parameter which can lead to
$w_{\circ }\sim -1.17$ and allows for quintom behavior. However, in
this paper, we incorporate GB dark energy density with a SCG
component without adding any matter content. In addition,
corporation GB or MGB with different CG models (\emph{i.e.} SCG and
GCG) would be help full in order to obtain exact solution for scalar
field and potential and would relieve us in order to determine some
bounds for free parameters of models. So considering the
cosmological solution for different compositions of GB and CG models
could show the importance of each one. Also, we would succeed in the
frame work where $\kappa ^{2}=8\pi G=M_{p}^{-2}=1$ and in the
natural unit where $(\hbar=c =1)$.

The outline of this paper is as follows: In next section, we
introduce the GB dark energy and calculate the deceleration and EoS
parameters. Then, in subsections 2.1 , 2.2 and 2.3 we investigate
corporation GB with SCG and GCG, in turn and then scalar field and
scalar potential are obtained by exact solution. In section III, the
same procedure has done for MGB energy density. In section IV, we
would investigate Adiabatic Sound Speed, $v^{2}$, which is one of
the critical physical quantity in the theory of linear perturbation.
In section V, we discuss on behavior of scalar field, scalar
potential and deceleration parameter \emph{versus} $x$ for GB and
MGB models and we gain some bounds for free parameters of models.
Finally, we summarize our results in Sec. VI.

\section{Gauss-Bonnet Dark energy in a flat universe}
The energy density of GB-DE is given by
\begin{equation}
\rho _{D}=\alpha \mathcal{G}  \label{1-1}
\end{equation}%
where $\alpha $ is a positive dimensionless parameter \cite{GRANDA}.
Gauss-Bonnet invariant $\mathcal{G}$ is topological invariant in
four dimensions and may lead to some interesting cosmological
effects in higher dimensional brane-world (for a review, see
\cite{nojirii}). It is defined as
\begin{equation}
\mathcal{G}=R^{2}-4R_{\mu \nu }R^{\mu \nu }+R_{\mu \nu \eta \gamma
}R^{\mu \nu \eta \gamma }  \label{1-2}
\end{equation}%
where $R$, $R_{\mu \nu }$ and $R_{\mu \nu \eta \gamma }$ are scalar
curvature, Ricci curvature tensor and Riemann curvature tensor,
respectively. In a  spatially flat FRW universe
\begin{equation}
d^{2}s=-dt^{2}+a^{2}\left( t\right) \left[ dr^{2}+r^{2}d\theta ^{2}
+r^{2}\sin ^{2}\theta d\phi^{2} \right]   \label{1-3}
\end{equation}%
the Eq. (\ref{1-1}) takes the form
\begin{equation}
\rho _{D}=24\alpha H^{2}\left( H^{2}+{\dot{H}}\right). \label{1-4}
\end{equation}%
By using the energy density $\rho _{D}$, without any matter
component, the Friedmann equation in flat universe in reduced Planck
mass unit ($8\pi G=\hbar=c=1$) is
\begin{equation}
H^{2}=\frac{1}{3}\rho _{D}=8\alpha H^{2}\left( H^{2}+{\dot{H}}
\right).   \label{1-5}
\end{equation}%
Defining the e-folding $x$\ with definition $x=lna=-ln(1+z)$, where
$z$\ is the redshift parameter and using $d/d(x)=\frac{1}{H}d/d(t)$,
we get the following differential equation
\begin{equation}
H^{2}+\frac{1}{2}\frac{dH^{2}}{dx}-\frac{1}{8\alpha }=0,
\label{1-6}
\end{equation}%
which immediately gives the solution
\begin{equation}
H( x) =\sqrt{\frac{1}{8\alpha }(1+\xi e^{-2x})}. \label{1-8}
\end{equation}%
The parameter $\xi$ is a constant of integration which is obtained
by $\xi =8\alpha H_{0}^{2}-1$. Also it gives $\alpha=(1+\xi)/(8
H_0^2)$. Using the continuity equation
\begin{equation}
\overset{\cdot }{\rho _{D}}+3H\left( 1+w_{D}\right) \rho _{D}=0
\label{1-9}
\end{equation}%
and Eqs. (\ref{1-4}),(\ref{1-5}), the equation of state (EoS)
parameter yields
\begin{equation}
w_{D}=-1-\frac{\dot{\rho_D}}{3H\rho_{D}}=-1-\frac{2}{3}\frac{\dot{H}}{H^{2}}=-1-\frac{2}{3}(\frac{1}{8\alpha
H^2}-1). \label{1-11}
\end{equation}
It is more preferable to write above equation in term of e-folding,
$x$. Hence, by using Eq. (\ref{1-8}), the EoS parameter can be
rewritten as
\begin{equation}
w_{D}=-1+\frac{2}{3}\left( \frac{\xi e^{-2x}}{1+\xi e^{-2x}}\right).
\label{1-12}
\end{equation}%
We see that the constant $\xi $ plays a crucial role in the behavior
of the EoS parameter. For $ \xi =0$ (\emph{i.e.} $8\alpha
H_{0}^{2}=1$), the EoS parameter for $\Lambda CDM$ model
($w_{\Lambda }=-1$) is retrieved. For $\xi >0$ the expanding
universe accelerates in quintessence phase ($-1<w_{D}<-1/3$). Using
Eqs. (\ref{1-5}) and (\ref{1-8}), the deceleration parameter is
calculated as
\begin{equation}
q=-1-\frac{{\dot {H}}}{H^{2}}=-\frac{1}{8\alpha
H^{2}}=-\frac{1}{1+\xi e^{-2x} }. \label{1-13}
\end{equation}
Since $\alpha $ and $H_{0}^{2}$ are positive parameters, so $\xi $
always must be greater than $-1$. Therefore, the deceleration
parameter is always negative except for $-1<\xi <0$. In this way,
the universe which is characterize by GB dark energy model could not
exhibit a transition from deceleration to acceleration phase for
$\xi\geq0 $, against what we expect from observations.

\subsection{Gauss Bonnet Standard Chaplygin Gas}\label{GB-SCG}
\begin{figure}[tbp]
\epsfxsize=7cm\centerline{\epsffile{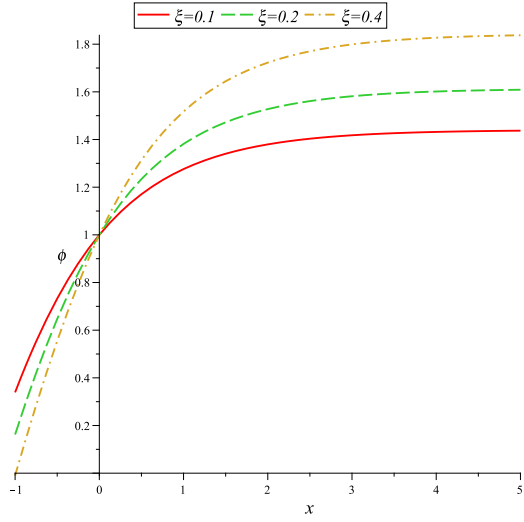}}
\bigskip
\caption{behavior of normalized scalar field \emph{versus} e-folding
$x$ for some values of $\xi \geq 0$ } \label{fig11}
\end{figure}

The SCG is a perfect fluid with an equation of state as
\begin{equation}
p_{SCG}=-\frac{A}{\rho },  \label{2-1}
\end{equation}%
where $p$, $\rho $ and $A$ are pressure, energy density and a
positive constant respectively. By substituting Eq.(\ref{2-1}) into
the continuity equation (\ref{1-9}), the energy density immediately
solved
\begin{equation}
\rho_{SCG} =\sqrt{A+Be^{-6x}},  \label{2-2}
\end{equation}%
where $B$ is an integration constant \cite{khodam2641}. Using the
standard scalar field DE model in which the energy density and
pressure are defined as
\begin{eqnarray}
\rho _{\phi } &=&\frac{1}{2}{\dot {\phi }}^{2}+V\left( \phi \right)
= \sqrt{A+Be^{-6x}},  \label{2-3} \\
p_{\phi } &=&\frac{1}{2}{\dot {\phi} }^{2}-V\left( \phi \right)
=\frac{-A}{\sqrt{A+Be^{-6x}}}, \label{2-4}
\end{eqnarray}%
and equating $p_{SCG}=p_{\phi }$ and $\rho _{SCG }=\rho _{\phi }$,
the scalar potential and kinetic energy term of SCG model are given
as
\begin{equation}
V\left( \phi \right) =\frac{2A+Be^{-6x}}{2\sqrt{A+Be^{-6x}}}
\label{2-5}
\end{equation}%
\begin{equation}
{\dot {\phi }}^{2}=\frac{Be^{-6x}}{\sqrt{A+Be^{-6x}}}. \label{2-6}
\end{equation}%
Also the EOS parameter becomes
\begin{equation}
w_{SCG}=\frac{p}{\rho }=-\frac{A}{A+Be^{-6x}}  \label{2-8}
\end{equation}
Equating the energy densities (\emph{i.e.}, $\rho_{SCG}=\rho_{D}$)
and EoS parameters (\emph{i.e.}, $w_{SCG}=w_{D}$), after using the
Friedmann equation (\ref{1-5}), constants $A$ and $B$ immediately
given by
\begin{eqnarray}
A &=& \frac{3}{(8\alpha )^{2}}\left[ (2+\xi e^{-2x})^{2}-1\right],
\label{2-9} \\
B &=& e^{6x}\left[ \left( \frac{3}{8\alpha }(1+\xi
e^{-2x})\right) ^{2}-A\right],   \label{2-7}
\end{eqnarray}
and hence the scalar potential and kinetic energy term rewritten as
\begin{eqnarray}
V\left(x\right) &=& \frac{1}{8\alpha}\left( 3+2\xi
e^{-2x}\right)=\frac{H_0^2}{1+\xi}\left( 3+2\xi e^{-2x}\right),
\label{2-10}\\
\dot {\phi} &=& \frac{1}{2}\sqrt{\frac{\xi e^{-2x}}{\alpha }}.
\label{2-11}
\end{eqnarray}
By inserting $\phi ^{\prime }=\dot {\phi}/H$ , where prime means
derivative with respect to $x=\ln a$, the differential equation
(\ref{2-11}) gives the normalized scalar field ($\phi =1$ at
present, $x=0$) in terms of $x$ as
\begin{equation}
\phi =1-\frac{\sqrt{2}}{2}\ln\left( \frac{1+2\xi e^{-2x}+2\sqrt{\xi
e^{-2x}(1+\xi e^{-2x})}}{1+2\xi +2\sqrt{\xi(1+\xi)}}\right).
\label{2-14}
\end{equation}%
It is easy to see that from Eq. (\ref{1-8}), at present, we must
have $1+\xi \geq 0$ and from (\ref{2-14}), it must be required that
$\xi(1+\xi) \geq 0$. Therefore in this model, we must have $\xi \geq
0$. As it is shown in Fig. \ref{fig11}, the normalized scalar field
grows up to a saturated value at late time in such a way that this
value exceeds for larger values of $\xi$. Also Eq. (\ref{2-10})
shows that the universe goes to a stable equilibrium at infinity
where $V(\infty)=3H_0^2/(1+\xi)$ and from (\ref{2-14}), the scalar
field reaches to $\phi(\infty)=1+(\sqrt{2}/2)\ln\left(1+2\xi
+2\sqrt{\xi(1+\xi)}\right).$

\subsection{Gauss Bonnet Generalized Chaplygin Gas}
The equation of state of generalized Chaplygin gas (GCG) defined as
\cite{bento4}
\begin{equation}
p=-\frac{A}{\rho ^{\delta -1}},  \label{3-1}
\end{equation}%
where $A$ is a constant and $1\leq \delta \leq 2$. For $\delta =2$,
it reaches to SCG model. The energy density, similar to previous
case, is given by
\begin{equation} \rho_{GCG} =\left( A+Be^{\left( -3\delta x\right)
}\right) ^{\frac{1}{\delta }}, \label{3-2}
\end{equation}%
and the scalar field model gives energy density and pressure of GCG
as
\begin{equation}
\rho _{\phi }=\frac{1}{2}{\dot {\phi} }^{2}+V\left( \phi \right)
=\left( A+Be^{-3\delta x}\right) ^{\frac{1}{\delta }}, \label{3-3}
\end{equation}
\begin{equation}
p_{\phi }=\frac{1}{2}{\dot {\phi} }^{2}-V\left( \phi \right)
=-A\left( A+Be^{-3\delta x}\right) ^{-\frac{\delta -1}{\delta }}.
\label{3-4}
\end{equation}%
After forward calculation, three quantities: the scalar potential,
kinetic term and EoS parameter are given by
\begin{equation}
V\left( x\right) =\frac{2A+Be^{-3\delta x}}{2\left( A+Be^{-3\delta
x}\right) ^{\frac{\delta -1}{\delta }}},  \label{3-5}
\end{equation}%
\begin{equation}
{\dot {\phi} }^{2}=\frac{Be^{-3\delta x}}{\left( A+Be^{-3\delta
x}\right) ^{\frac{\delta -1}{\delta }}}, \label{3-7}
\end{equation}%
\begin{equation}
w_{GCG}=\frac{p}{\rho }=-\frac{A}{A+Be^{-3\delta x}}.  \label{1-10}
\end{equation}%
Also same as previous, the constants $A$ and $B$ reconstructed as
\begin{eqnarray}
A&=&\frac{3+\xi e^{-2x}}{\left( 8\alpha \right) ^{\delta }}\left[
3\left( 1+\xi e^{\left( -2x\right) }\right) \right] ^{\delta -1},
\label{3-10} \\
B&=&e^{3\delta x}\left[ \left( \frac{3}{8\alpha }(1+\xi
e^{-2x})\right) ^{\delta }-A\right]  \label{3-8}
\end{eqnarray}%
and the potential and dynamics of GB-GCG can be written as
\begin{eqnarray}
V\left( x\right) &=&\frac{3+2\xi e^{-2x}}{8\alpha },  \label{3-11}
\\
{\dot {\phi} }&=&\frac{1}{2}\sqrt{\frac{\xi e^{-2x}}{\alpha }}.
\label{3-12}
\end{eqnarray}%
As it is seen, the potential and dynamics of GB-GCG are not a
function of parameter $\delta $ and are exactly similar to previous
case (see Eqs. (\ref{2-10}) and (\ref{2-11})). Therefore the
reconstructed scalar potential and scalar field obtained by previous
Eqs. (\ref{2-10}) and (\ref{2-14}). It is worthwhile to mention that
both models that we have been studied, encourage with an essential
problem. Despite of observational predictions, the phase transition
between deceleration to acceleration expansion did not happen in
GB-DE model. Therefore we will study on the MGB model, which may
alleviate this problem.

\section{Modified Gauss Bonnet Dark Energy}
The energy density MGB has been defined by
\begin{equation}
\rho _{D}=3H^{2}(\gamma H^{2}+\lambda \overset{\cdot }{H}),
\label{4-1}
\end{equation}%
where $\gamma $ and $\lambda $ are dimensionless constants
\cite{GRANDA}. The Friedmann equation in dark dominated flat
universe gives
\begin{equation}
\gamma H^{2}+\frac{1}{2}\lambda \left( \frac{dH^{2}}{dx}\right)
-1=0, \label{4-2}
\end{equation}%
and the Hubble parameter given by
\begin{equation}
H(x)=\sqrt{\frac{1}{\gamma }(1+\eta e^{-\frac{2\gamma x}{\lambda
}})}, \label{4-3}
\end{equation}
where $\eta$ is an integration constant which is obtained by $\eta
=\gamma H_{0}^{2}-1$. The EoS parameter becomes
\begin{equation}
w_{D}=-1-\frac{2}{3}\frac{\overset{\cdot }{H}}{H^{2}}=-1+\frac{2\gamma }{%
3\lambda }\left( \frac{\eta e^{-\frac{2\gamma x}{\lambda }}}{%
1+\eta e^{-\frac{2\gamma x}{\lambda }}}\right)  \label{4-4}
\end{equation}%
and deceleration parameter is obtained as follows
\begin{equation}
q=-1-\frac{\overset{\cdot }{H}}{H^{2}}=-1+\frac{\gamma }{\lambda
}\left(
\frac{\eta e^{-\frac{2\gamma x}{\lambda }}}{1+\eta e^{-\frac{%
2\gamma x}{\lambda }}}\right).  \label{4-21}
\end{equation}
For positive values of $\gamma$ and $\lambda $, from Eq.
(\ref{4-3}), it is easy to see that $\eta $ must be always greater
than $-1$ and from Eq. (\ref{4-21}), a transition from deceleration
to acceleration is expected provided that $\eta \geq 0$. Detailed
discussion were transferred to section \ref{discussion}.

\subsection{Modified Gauss Bonnet And SCG}
Now we want to investigate on the correspondence between MGB-SCG
models and reconstruct the potential and dynamics of scalar field.
Same as before, equating energy densities (\emph{i.e.}, Eqs.
(\ref{2-2} ) and (\ref{4-1})) and EoS parameters (\emph{i.e.},
(\ref{2-8}) and (\ref{4-4})), yield
\begin{eqnarray}
A&=&\frac{9}{\gamma ^{2}}\left( 1+\eta e^{-\frac{2\gamma x}{\lambda }%
}\right) \left[ 1+(1-\frac{2\gamma }{3\lambda })\eta e^{
-\frac{2\gamma x}{\lambda }}\right],   \label{4-7}\\
B&=&e^{6x}\left[ \left( \frac{3}{\gamma }(1+\eta e^{-\frac{2\gamma x}{%
\lambda }})\right) ^{2}-A\right].  \label{4-5}
\end{eqnarray}%
By substituting $A$ and $B$ in Eqs. (\ref{2-5}) and (\ref{2-6}), we
find
\begin{eqnarray}
V\left( x\right) &=&\frac{3}{\gamma }\left[ 1+(1-\frac{\gamma }{3\lambda }%
)\eta e^{-\frac{2\gamma x}{\lambda }}\right], \label{4-8}\\
\dot\phi&=&\sqrt{\frac{2\eta }{\lambda }e^{-\frac{%
2\gamma x}{\lambda }}} , \label{4-9}
\end{eqnarray}%
which immediately gives the normalized scalar field as
\begin{equation}
\phi =1-\frac{\sqrt{2}}{2}\sqrt{\frac{\gamma}{\lambda}}\ln\left(
\frac{1+2\eta e^{-\frac{2\gamma x}{\lambda }}+2\sqrt{\eta
e^{-\frac{2\gamma x}{\lambda }}(1+\eta e^{-\frac{2\gamma x}{\lambda
}})}}{1+2\eta +2\sqrt{\eta(1+\eta)}}\right). \label{4-10}
\end{equation}%
The behavior of scalar field in this model is the similar to GB-DE
model as discussed in Sec. \ref{GB-SCG}.

\subsection{Modified Gauss Bonnet And GCG}
As previous, the constants $A$ and $B$ are
\begin{eqnarray}
A&=&(\frac{3}{\gamma })^{\delta }\left( 1+\eta e^{^{-\frac{2\gamma x}{\lambda }%
}}\right) ^{\delta -1}\left[ 1+\left( 1-\frac{2\gamma }{3\lambda
}\right) \eta e^{-\frac{2\gamma x}{\lambda }}\right],   \label{5-3}\\
B&=&e^{3\delta x}\left[ \left( \frac{3}{\gamma }(1+\eta e^{-\frac{2\gamma x}{%
\lambda }}\right) ^{\delta }-A\right].   \label{5-1}
\end{eqnarray}
and the potential and dynamics of MGB-GCG are given by
\begin{eqnarray}
V(\phi )&=&\frac{3}{\gamma }\left[ 1+\left( 1-\frac{\gamma
}{3\lambda }\right) \eta e^{^{-\frac{2\gamma x}{\lambda }}}\right]
\label{5-4}\\
\dot\phi&=&\sqrt{\frac{2\eta }{\lambda }e^{^{-\frac{2\gamma x}{%
\lambda }}}} \label{5-5}
\end{eqnarray}
which are exactly similar to (\ref{4-8}) and (\ref{4-9}) in previous
model. Therefore the behavior of normalized scalar field and
potential are the same as MGB-SCG model.

\section{Adiabatic Sound Speed}
Investigation of the squared of sound speed, $v^{2}$, would help us
to determine the growth of perturbation in linear theory
\cite{peebleratra}. The sign of $v_{s}^{2}$ plays a crucial role in
determining the stability of the background evolution. Positive sign
of $v^{2}$ shows the periodic propagating mode for a density
perturbation and probably represents an stable universe against
perturbations. The negative sign of it shows an exponentially
growing/decaying mode in density perturbation, and can show sounds
of instability for a given model. The squared of sound speed is
defined as \cite{peebleratra}
\begin{equation}
v^{2}=\frac{dP}{d\rho }=\frac{\overset{\cdot }{P}}{\overset{\cdot
}{\rho }} \label{11-1}
\end{equation}
In a dark dominated flat universe, it can be written as
\begin{equation}
v^{2}=-1-\frac{1}{3}\left( \frac{\overset{\cdot \cdot }{H}}{\overset{\cdot }{%
H}H}\right).   \label{11-3}
\end{equation}%
and it immediately gives a constant squared of sound speed for GB-DE
as $v^{2}=-1/3$. Therefore it may reveal an instability against the
density perturbation in GB-DE model. For MGB-DE, Eq.(\ref{11-3})
gives
\begin{equation}
v^{2}=-1+\frac{2\gamma }{3\lambda }.  \label{11-4}
\end{equation}
It shows that $v^{2}$ can be positive provided that
$\gamma/\lambda>3/2$. Thus an stable DE dominated universe may be
achieved in this model. In the next section we would improve this
bound for $\gamma/\lambda$ in a proper way.

\section{\protect\bigskip Discussion}\label{discussion}
\begin{figure}[tbp]
\epsfxsize=7cm\centerline{\epsffile{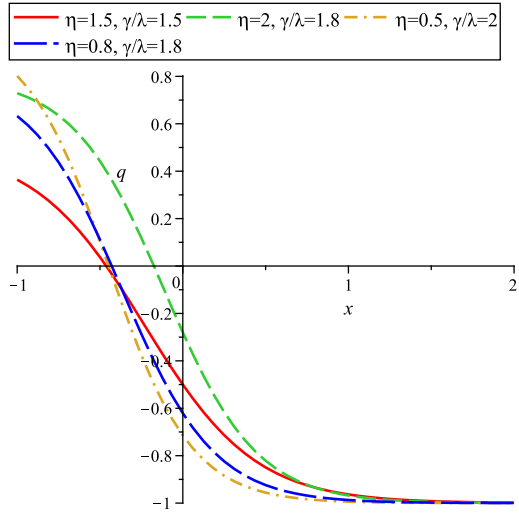}}
\bigskip
 \caption{\small{The behavior of
deceleration parameter $q$ \emph{versus} e-folding $x$ for various
$\eta$ and $\gamma/\lambda$. The transition from deceleration to
acceleration was happened around $x\sim 0.5$} } \label{fig1}
\end{figure}

\begin{figure}[tbp]
\epsfxsize=7cm\centerline{\epsffile{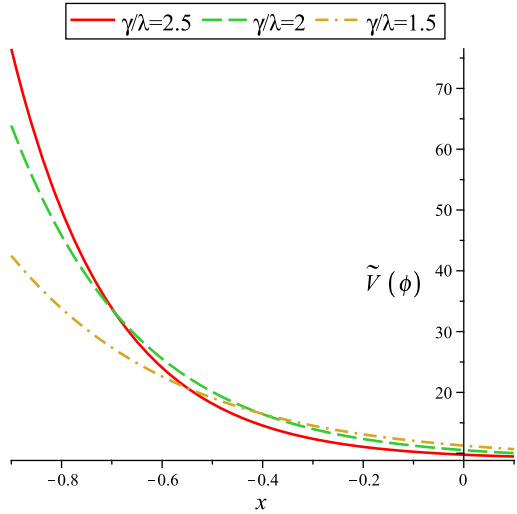}}
\bigskip
\caption{\small{behavior of $\protect\overset{\sim }{V(\protect\phi
)}$ \emph{versus} e-folding $x$ for various $\gamma/\lambda$ and
$\eta=1.5$ } }\label{fig2}
\end{figure}

We are interesting to focus on MGB-DE model. At first, we start with
Eq. (\ref{4-21}) and plot the deceleration parameter with respect to
$x$ in Fig. \ref{fig1}. It shows that the deceleration parameter
transits from deceleration ($q>0$) to acceleration ($q<0$) in some
point at the past. The parameters $\eta $ and $\gamma/\lambda$ play
a crucial rule for this point. As $\eta $ or $\gamma/\lambda$ adopt
bigger values, the transition point approaches to present time. By
choosing the best values of $q_{0}~(\sim -0.6)$ and inflection point
as $(x\simeq -0.5)$ which has been parameterized recently
\cite{pavon et al, kh-ka, daly}, we obtain some bounds for $\eta $
and $\gamma/\lambda$ as follow
\begin{equation}
0<\eta <2.5~~~~~~ 1.5\leq \frac{\gamma }{\lambda }\leq 3
\end{equation}
Using Eq. (\ref{4-8}) for MGB model, we plot $\overset{\sim }{V(\phi
)}=\gamma V(\phi )$ \emph{versus} $x$ for different values of
$\gamma/\lambda$ and $\eta =1.5$ in Fig. \ref{fig2}. This figure
shows that as time goes, $\overset{\sim }{V(\phi )}$ is decreasing
to small values and the potential will reach to a constant at
infinity. In addition, by increasing the ratio of $\gamma/\lambda$,
the tracking potential adopts bigger values at future.

As it is seen, the potential describes a tracker solution. According
to the quintessential tracker solution, our universe undergoes a
phase from $w=0$ to $w=-1$ and the effective EoS is $w_{eff}=-0.75$
\cite{steinhardt}. The huge advantage of the tracker solution is
that it allows the quintessence model to be insensitive to initial
conditions \cite{YOO}. So we use this feature in order to improve
obtained bounds of parameters. In this way, Eqs. (\ref{4-4}) and
(\ref{5-4}), for matter dominated universe ($w=0$), leads to $
V(\phi )=3/(2\gamma -3\lambda )$. On the other hand the quantity $
V(\phi )$ for quintessence barrier ($w=-1/3$) reach to $V(\phi )=2/(
\gamma -\lambda )$, so that the value $\gamma/\lambda=1$ is illegal.
It is also consistent with what we got from investigation of the
deceleration parameter. Finally, the potential might give a tracking
solution provided that $1.5<\gamma/\lambda\leq 3$.

\section{conclusion}
In this paper, the reconstruction of GB-DE and some variety of
Chaplygin gas have been studied. We obtained exact solutions for
reconstructed scalar field and its potential in each models (GB-SCH,
GB-MCG, MGB-SCG and MGB-MCG). According to cosmological predictions
and historical evolutions, some models should be rejected
(\emph{i.e.}, models combined with GB-DE) and another models which
have been combined with MGB-DE can be permitted to express the
evolution of the universe. The equation of state and deceleration
parameters for both GB and MGB models were calculated. In GB-DE
model, the deceleration parameter was always negative except for
$-1<\xi <0$. This fact was shown that a transition from deceleration
to acceleration expansion could not have happened in the past that
is contrary to the facts of cosmology. Also it was easily shown that
the EoS parameter in GB-DE model would not ever reach to phantom
phase (\emph{i.e.} $w_D<-1$). We showed that for $\xi =0$
(\emph{i.e.} $8\alpha H_{0}^{2}=1$), the EoS parameter for
$\Lambda$CDM model was retrieved. Investigation on the squared of
sound speed, revealed an instability of model against density
perturbation in GB-DE model.

In MGB-DE model, we found that the transition from deceleration to
acceleration is permitted just for a limited range of values of
$\eta $ and $\gamma/\lambda $. Choosing the best values for
deceleration parameter at present and deflection point, according to
observations, some bounds of $0<\eta <2.5$ and $ 1.5\leq
\gamma/\lambda\leq 3$ were obtained. We showed that by redefining
$\gamma V(\phi )=\overset{\sim }{ V(\phi )}$, the scalar potential
decreased to smaller values and will reach to a saturated constant
at late time. Our investigation on $ V(\phi )$ for two phases,
matter dominate and quintessence, showed that $\gamma/\lambda$,
could not take two values $1$ and $3/2$.

It will be interesting to find the constraints of these models
against the data of cosmological observations and structure
formation. We hope to discuss these issues in the future.


\begin{thebibliography}{99}

\bibitem{A.J.Riess} A. G. Riess \emph{et al}., Astron. J. \textbf{116}, 1009
(1998).

\bibitem{S.Perlmutter} S. Perlmutter \emph{et al}.,Nature \textbf{391, }51 (1998).

\bibitem{M. Hicken} M. Hicken \emph{et al}., Astrophys. J. \textbf{700},
1097,(2009).

\bibitem{Tegmark} M. Tegmark \emph{et al}., Astrophys. J.\textbf{606}, 702 (2004).

\bibitem{K. Abazajian} K. Abazajian \emph{et al}., [SDSS Collaboration] Astron. J.
\textbf{129}, 1755 (2005).

\bibitem{spergel} D.N. Spergel \emph{et al}., Astrophys. J. Suppl. \textbf{148},
175 (2003).

\bibitem{komatsu} E. Komatsu \emph{et al}., [WMAP Collaboration], Astrophys. J.
Suppl. \textbf{180}, 330 (2009).

\bibitem{modG1} S. Nojiri and S.D. odintsov, Int. j. Geom. Methods M.
\textbf{4}, 115 (2007).
\bibitem{modG2} I. Martino, M. Laurentis and S. Capozziello,
Universe, \textbf{1}, 199, 2015.
\bibitem{modG3} S. Capozziello \emph{et al}., Universe,
\textbf{1}, 199 (2015).
\bibitem{modG4} S. Bahamonde, C.G. B\"{o}hmer, F.S.N. Lobo
and D. S\'{a}ez-G\'{o}mez, Universe, \textbf{1}, 186 (2015).
\bibitem{modG5} S. Basilakos, N.E. Mavromatos and J. Sol\`{a}, Universe, \textbf{2},
14 (2016).
\bibitem{modG6} L. Iorio \emph{et al}., Physics of the Dark Universe, \textbf{13},
111 (2016).
\bibitem{modG7} Lorenzo Iorio, Ninfa Radicella and Matteo Luca Ruggiero,
JCAP, \textbf{08}, 21 (2015).
\bibitem{modG8} K. Rezazadeh, A. Abdolmaleki and K.
Karami, JHEP, \textbf{01}, 131 (2016).
\bibitem{modG9} K. Karami, A. Abdolmaleki, S.
Asadzadeh and Z. Safari, Eur. Phys. J. C \textbf{73}, 2565 (2013).
\bibitem{modG10} A. Khodam-Mohammadi, P. Majari and M. Malekjani, Astrophys. Space
Sci. \textbf{331}, 673 (2011).
\bibitem{modG11} A. De Felice and S. Tsujikawa, living
Rev. relativ. \textbf{13}, 3 (2010).
\bibitem{modG12} T.P. Sotiriou and V. Faraoni,
Rev. Modern Phys. \textbf{82}, 451 (2010).
\bibitem{modG13} A. Zanzi, Universe \textbf{1}, 446 (2015).
\bibitem{modG14} Yi-Fu Cai, S. Capozziello, M. De Laurentis and E.N. Saridakis,
Rept. Prog. Phys. \textbf{79}, 106901 (2016).

\bibitem{Chavanis} P.H. Chavanis, Universe, \textbf{1}, 357 (2015).

\bibitem{sola1409} A. Gomez-Valent \emph{et al}., JCAP \textbf{01} 004(2015).

\bibitem{sola1412} A. Gomez-Valent \emph{et al}., Mon. Not .Roy. Astron. Soc.
\textbf{448}, 2810 (2015).

\bibitem{sola31} J.A.S. Lima, S. Basilakos, and J. Sol\`{a}, Mon. Not. Roy. Astron. Soc. \textbf{431}, 923
(2013).

\bibitem{Perico} E.L.D. Perico, J.A.S. Lima, S. Basilakos, and J. Sola, Phys.
Rev. D \textbf{88}, 063531 (2013).

\bibitem{Basilakos13} S. Basilakos, J. A. S. Lima, and J. Sola, Int. J. Mod. Phys. D
\textbf{22}, 1342008 (2013).

\bibitem{bleem} L.E. Bleem \emph{et al}., Astrophys. J. \textbf{216}, 27
(2015).

\bibitem{Lima} J.A.S. Lima, M. Trodden, Phys. Rev. D \textbf{53}, 4280(1996).

\bibitem{GRANDA} L.N. Granda, Mod. Phys. Lett. A \textbf{28}, 1350117 (2013).

\bibitem{kofinas} G. Kofinas, R. Maartens and E. Papantonopoulos, JHEP \textbf{0310}, 066
(2003).

\bibitem{brown} R.A. Brown, R. Maartens, E. Papantonopoulos and V.
Zamarias, JCAP \textbf{0511}, 008 (2005).

\bibitem{nojiri222} S. Nojiri, S.D. Odintsov and M. Sasaki, Phys. Rev. D \textbf{71}, 123509  (2005).

\bibitem{chaplygin} S. Chaplygin, Sci. Mem. Moscow Univ. Math. Phys. \textbf{21}, 1
(1904).

\bibitem{kamens} A.Y. Kamenshchik, U. Moschella, and V. Pasquier, Phys.
Lett. B \textbf{ 511}, 265 (2001).

\bibitem{bento} M.C. Bento, O. Bertolami, and A.A. Sen, Phys. Rev. D \textbf{%
66}, 043507 (2002).

\bibitem{gorini} V. Gorini, A. Kamenshchik, U. Moschella, V. Pasquier;
[arXiv:gr-qc/0403062].

\bibitem{zhu} Z.H. Zhu, Astron. Astrophys., \textbf{423}, 421 (2004).

\bibitem{bento2} M.C. Bento, O. Bertolami and A.A. Sen , Phys. Lett.
B \textbf{575}, 172 (2003).

\bibitem{bilic} N. Bilic, G.B.Tupper and R.D. Viollier, Phys.
Lett. B \textbf{535}, 17 (2001).

\bibitem{denbath} U. Debnath, A. Banerjee, and S. Chakraborty, Class.
Quantum Grav. \textbf{21}, 5609 (2004).

\bibitem{Brown1} R.A. Brown, Gen. Rel. Grav. \textbf{39}, 477 (2007).

\bibitem{Cai1} R.G. Cai, H.S. Zhang and A. Wang, Commun. Theor. Phys.
\textbf{44}, 948 (2005).

\bibitem{nojirii} S. Nojiri, S. D. Odintsov, and S. Ogushi, Int. J. Mod. Phy. A \textbf{17}, 4809 (2002).

\bibitem{khodam2641} M. Malekjani, A. Khodam-mohammadi, Int. J. Mod. Phys. D \textbf{20}, 281 (2011).

\bibitem{bento4} M.C. Bento, O. Bertolami and A.A. Sen, Phys.Rev. D \textbf{70}, 083519 (2004) .

\bibitem{peebleratra} P.J.E. Peebles and B. Ratra, Rev. Mod. Phys. \textbf{75}, 559
(2003).

\bibitem{pavon et al} D. Pavon \emph{et al}., Phys. Rev. D \textbf{ 86}, 083509 (2012).

\bibitem{kh-ka} A. Khodam-mohammadi, E. Karimkhani, Int. J. Mod. Phys. D \textbf{23}, 1450081 (2014).

\bibitem{daly} R.A. Daly \emph{et al}., Astrophys. J. \textbf{677}, 1 (2008).

\bibitem{steinhardt} P.J. Steinhardt \emph{et al}., Phys.Rev. D \textbf{59}, 123504
(1999).

\bibitem{YOO} J. Yoo and Y. Watanabe, Int. J. Mod. Phys. D \textbf{21}, 1230002
(2012).
\end{thebibliography}
\end{document}